\documentclass{aa}
\usepackage{graphicx}
\usepackage{txfonts}
\usepackage{breqn}
\usepackage{color}

\def \gray {$\gamma$-ray\xspace}
\def \grid {AGILE-\textit{GRID}\xspace}

\begin{document}

   \title{A new gamma-ray source unveiled by AGILE in the region of Orion}

   \author{N. Marchili \inst{1} \and
           G. Piano \inst{1} \and
           M. Cardillo \inst{1} \and
           A.  Giuliani \inst{2}
           S. Molinari \inst{1} \and
           M. Tavani  \inst{1}
          }

   \institute{INAF - IAPS, Via Fosso del Cavaliere 100, 00133, Roma, Italy\\
              \email{nicola.marchili@gmail.com}
         \and
             INAF - IASF Milano, Via E. Bassini 15, I-20133, Milano, Italy\\
             }

   \date{Received ...; accepted ...}

  \abstract
   {Diffuse galactic \gray emission is produced by the interaction of cosmic rays (CRs) with the interstellar environment. The study of \gray emission is therefore a powerful tool to investigate the origin of CRs and the processes through which they are accelerated.}
   {We aim to gain deeper insights of the nature of \gray emission in the region of Orion, which is one of the best studied sites of on-going star formation, by analysing data from the AGILE satellite. Because of the large amount of interstellar medium (ISM) present in it, the diffuse \gray emission expected from the Orion region is relatively high. Its separation from the galactic plane also ensures a very low contribution from foreground or background emission, which makes it an ideal site for studying the processes of particle acceleration in star forming environments.}
   {The AGILE data are modelled through a template that quantifies the \gray diffuse emission expected from atomic and molecular hydrogen. Other sources of emission, such as inverse Compton (IC) scattering on interstellar radiation fields (ISRF) and extragalactic background, can be modelled as an isotropic contribution.}
   {\gray emission exceeding the amount expected by the diffuse emission model is detected with high level of significance. The main excess is in the high-longitude part of Orion A, which confirms previous results from the $Fermi$ Large Area Telescope. A thorough analysis of this feature suggests a connection between the observed \gray emission and the B0.5 Ia star $\kappa$ Orionis.}
   {We present the results of the investigation of \gray diffuse galactic emission from the region of Orion. The comparison between modelled and observed emission points towards the existence of larger-than-expected \gray flux from a $1^\circ$ radius region centred in $\kappa$ Orionis, compatible with the site where stellar wind collides with the ISM. Both scattering on dark gas and cosmic-ray acceleration at the shock between the two environments are discussed as possible explanations, with the latter hypothesis being supported by the hardness of the energy spectrum of the emission. If confirmed, this would be the first direct detection of \gray emission from the interaction between ISM and a single star's stellar wind.}


   \maketitle
%
%-------------------------------------------------------------------

\section{Introduction}

More than a century after the discovery of cosmic rays (CRs), the huge effort put into the investigation of their origin and the mechanisms of their acceleration provided us with a basic picture that still leaves open some fundamental questions.

Convincing arguments support the thesis that shocks of young supernova remnants are the main site of CR acceleration in the Galaxy \citep{Blandford87}. Their isotopic and elemental abundances suggest that CR nuclei are a mixture of $\sim20$\% of ejecta from high-mass stars and $\sim80$\% of gas from interstellar medium (ISM; \citealt{Binns05}; \citealt{Rauch09}).
The contribution of high-mass stars to the composition of CRs is therefore substantial; it is less clear, instead, how important their contribution is to the acceleration of particles. High-mass stars tend to cluster in large associations hosted by molecular clouds. Here, the colliding winds may create shock waves. Diffusive shock acceleration has been invoked as the most likely process for the acceleration of CRs \citep{Berezhko99}; in the case of high-mass star clusters, CR acceleration could take place as a consequence of the interaction among stellar winds \citep{Bykov01}. There is strong evidence supporting the hypothesis that the \gray emission detected by the $Fermi$ Large Area Telescope (LAT) in the Cygnus Superbubble is due to freshly accelerated CRs within the cavities carved by stellar winds from high-mass stars \citep{Ackermann11}. 

\cite{Munar-Adrover11} cross-checked the $Fermi$ First Source Catalog \citep{Abdo10} with catalogs of massive young stellar objects (MYSOs), Wolf-Rayet and Of-type stars, and OB associations, finding significant spatial coincidence between \gray sources and MYSOs. This indicates that CR acceleration could also take place around single stellar objects, when they form jets that could interact with the surrounding material. \cite{Casse80} hypothesised that acceleration of CRs could be triggered by OB stars too, in shocked regions at the boundary between ISM and the stellar wind.
\cite{Voelk82} investigated this case further reaching the conclusion that stellar winds might be sources of low energy CRs due to acceleration of nuclear particles, while higher energies could only be achieved through the contribution of electrons, provided that they can efficiently be injected and accelerated at the shock. The \gray luminosity from these regions is expected to be low.

In this paper, we present the analysis of \gray data concerning the diffuse emission towards Orion --- the nearest and most frequently observed site of on-going star formation (see \citealt{Bally08} and references therein) --- and neighbouring regions. The data were collected by \textit{GRID} (Gamma-Ray Imaging Detector, \citealt{Barbiellini2002}, \citealt{Prest2003}), which is the silicon-tracker imager on board the \gray satellite AGILE \citep{Tavani09}. Orion is embedded in the Orion-Eridanus superbubble; it comprises the stellar association Orion OB1, and the giant molecular clouds Orion A and Orion B. Given its importance for the study of massive star formation mechanisms and its relatively high distance from the galactic plane --- which reduces the contamination from foreground and background emission --- the region of Orion has been the target of intensive, high-resolution observations over a wide range of frequencies; in particular, because of its high concentration of molecular hydrogen, systematic observations have been performed in the mm to infrared bands, where CO abundances could be traced and turned into H$_2$ abundances using the X$_{\textrm{CO}}$ conversion factor. 

In the \gray band, the first detection of Orion belongs to COS-B \citep{Caraveo80}, followed by EGRET in the 90s \citep{Digel95}. The sensitivity and spatial resolution reached by $Fermi$ allowed a thorough study of the distribution of \gray emission from the whole region \citep{Ackermann12}. The detection of an excess of \gray emission from the high-longitude part of Orion A (l > $212^\circ$) was interpreted in terms of a nonlinear relation between H$_2$ and CO densities in diffuse molecular clouds. Our results confirm the findings of $Fermi$ about the excess of \gray emission; we propose however a different interpretation of its origin.

This paper is structured as follows: in Sec. 2 we describe the data, with the calibration procedures; in Sec. 3 we explain the modelling of the galactic diffuse \gray emission; the difference between the best fit model and the AGILE data is presented in Sec. 4, while in Sec. 5 we discuss the possible origin of the detected \gray excess. The summary is reported in Sec. 6.

%--------------------------------------------------------------------
\section{Observations and data calibration}

\begin{figure}
   \centering
   \includegraphics[width=0.96\columnwidth]{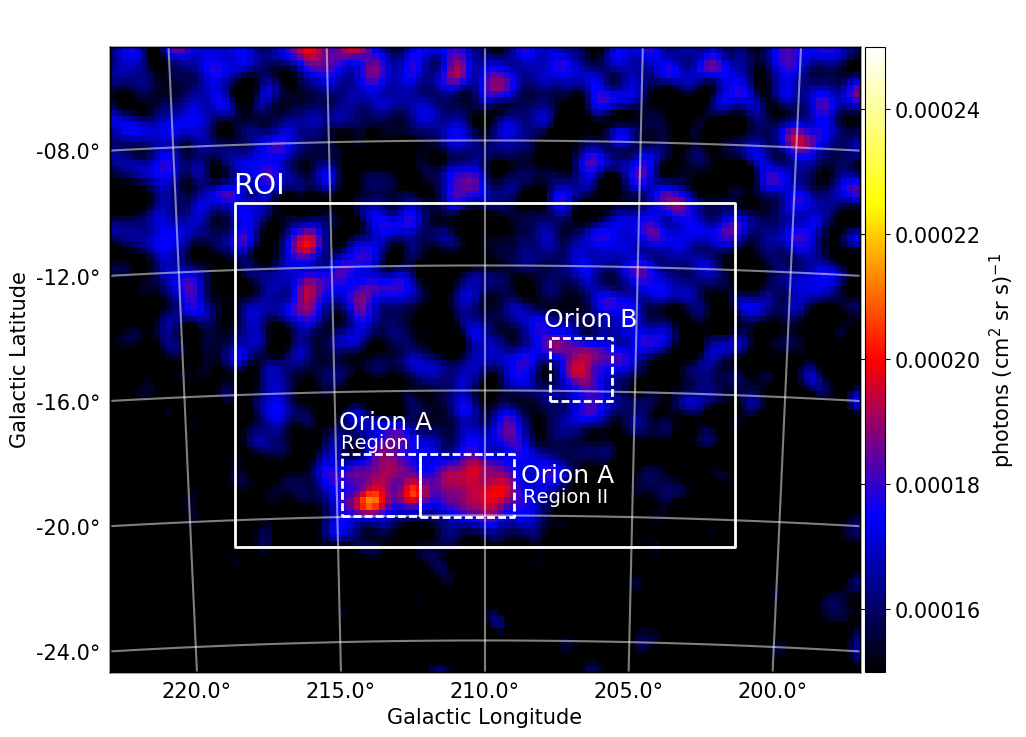}
      \caption{
      AGILE map of Orion. The large continuous rectangle highlights the region of interest, over which the best fit model was computed. The dashed rectangles, from left to right, show respectively the subregions 'Region I' and 'Region II' of Orion A, together with Orion B.
	}   
      \label{fig:mapA}
\end{figure}

We analysed \grid data in the area highlighted in Fig. \ref{fig:mapA} as the Region Of Interest (ROI), a $15^\circ\times11^\circ$ rectangle centred in {\it(l, b)}=[210.5, -15.5]. The data integrate observations performed by AGILE in the time span that goes from November 2009, when the satellite started operating in spinning mode, to March 2017. The \gray energy range to which \grid is sensitive goes from 30 MeV to 30 GeV. The point spread function (PSF) at 100 MeV is $4.2^{\circ}$, while at 400 MeV is $1.2^{\circ}$ ($68\%$ containment radius; see \citealt{Sabatini2015}).

\grid data have been analysed using the \verb+Build_22+ scientific software, while calibrated filter and response matrices were respectively \verb+FM3.119+ and \verb+I0025+. The standard configuration has been applied to the AGILE data analysis of incoming photons, which applies South Atlantic Anomaly event cuts and $80^{\circ}$ Earth albedo filtering; all incoming \gray photons with off-axis higher than $60^{\circ}$ have been discarded.

The AGILE multi-source likelihood analysis (MSLA) software \citep{Bulgarelli2012} provided the tools for the calculation of statistical significance and flux determination, for both point-like and extended sources. This software is based on the likelihood analysis developed by \cite{Mattox1996} for EGRET data, which assesses the reliability of a source detection by looking at the ratio between the maximum-likelihood values of two alternative hypotheses, namely a no-source model with another that postulates the existence of sources in the field.

Within the ROI, following the example of \cite{Ackermann12}, we defined the three subregions "Orion A, Region I", "Orion A, Region II", and "Orion B", which refer, respectively, to the giant molecular cloud Orion A and B, the former being divided into two separate zones. These are shown, as dashed rectangles, in Fig. \ref{fig:mapA}. In order to reduce the noise, a 3-point smoothing has been applied to the data.

%--------------------------------------------------------------------
\section{Data modelling}

\subsection{Galactic diffuse emission}

Following the approach described in \cite{Giuliani04}, the galactic \gray diffuse emission as a function of galactic coordinates ($l, b$), distance to the emitting medium ($r$), and photon energy ($E$), $g(l, b, r, E)$, was modelled as follows: 
\begin{dmath}
g(l, b, r, E) = [q_{\textrm{pp}}(l, b, r, E)+q_{\textrm{br}}(l, b, r, E)][n_{\textrm{HI}}(l, b, r)+2 \, n_{\textrm{H}_2}(l, b, r)]+q_{\textrm{IC}}(l, b, r, E) n_{\textrm{ph}}(l, b, r)
\end{dmath}
where $q_{\textrm{pp}}$ and $q_{\textrm{br}}$ are the \gray emissivities per hydrogen atom due to proton-proton scattering and Bremsstrahlung, respectively; $q_{\textrm{IC}}$ is the \gray emissivity per target photon due to Inverse-Compton (IC) scattering; $n_{\textrm{HI}}$, $n_{\textrm{H}_2}$ and $n_{\textrm{ph}}$ are the densities of atomic hydrogen, molecular hydrogen, and InterStellar Radiation Field (ISRF) photons.

The HI distribution in the galaxy was retrieved from the Leiden/Argentine/Bonn (LAB) 21-cm Survey (see \citealt{Kalberla05}; \citealt{Hartmann97}), while the H$_2$ column densities, which cannot be determined through direct observation, were assessed using the CO survey described by \cite{Dame01}; the CO has been converted into H$_2$ abundance using a CO-to-H$_2$ conversion factor, X$_{\textrm{CO}}$.

Due to the small variations of the ISRF across the region of Orion, the inverse-Compton contribution to the total \gray emission --- which in the ROI is anyway marginal --- was approximated as an isotropic component.

\begin{figure*}
   \centering
   \includegraphics[width=0.96\textwidth]{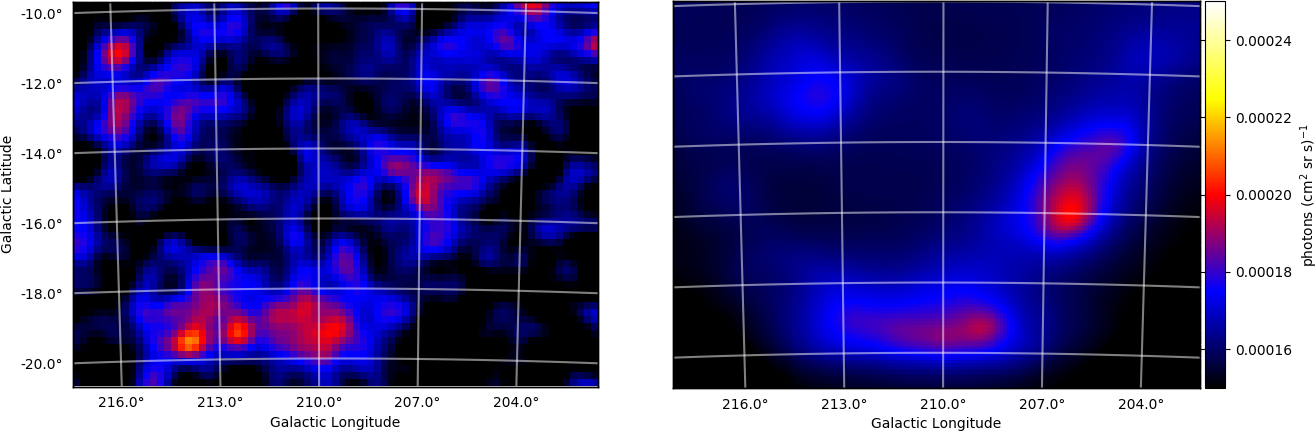}
      \caption{
      Comparison between the AGILE (left panel) and the modelled (right panel) maps of the Orion region.
	}   
      \label{fig:mapToModel}
\end{figure*}

The comparison between the modelled galactic \gray diffuse emission and the AGILE maps required the $g$ dependence on $r$ and $E$ to be removed. For this work, the whole sky was divided into squares of $0.1\times0.1$ deg$^2$; the $g(l, b)$ values we analysed were obtained by integrating $g(l, b, r, E)$ both over distance, and over the energy range that goes from 100 to 50000 MeV.

\subsection{Complete model}
Galactic diffuse emission is not the only source of \gray photons in the AGILE bandwidth. The \gray flux detected by AGILE also comprises point-like sources and isotropic extragalactic background. 
Given that the Second AGILE-GRID Source catalog doesn't contain any object in the ROI, the contribution from the former ones can be ignored. The extragalactic contribution, instead, can be modelled as an isotropic component.

A piece of software was developed to provide us with the best fit to the complete emission model presented above. It allows to separately treat the contributions that HI and H$_2$ provide to the observed \gray sky ($S_{\textrm{HI}}(l, b)$ and $S_{\textrm{H}_2}(l, b)$, respectively).

The total AGILE flux, $S_{\textrm{tot}}$, was fitted, either on the whole ROI or on specific sub-regions, as a linear combination of $S_{\textrm{HI}}$ and $S_{\textrm{H}_{2}}$:
\begin{equation}
S_{\textrm{tot}}(l, b) = \alpha \,S_{\textrm{HI}}(l, b)+\beta \,S_{\textrm{H}_2}(l, b)+\epsilon
\label{eq:fit}
\end{equation}
where the best fit proportionality factors $\alpha$ and $\beta$ were calculated, through minimisation of least squares, along with the isotropic component $\epsilon$ (which includes, as explained above, also the IC contribution). The results of this modelling procedure are displayed in Fig. \ref{fig:mapToModel}, where the AGILE map (left panel) is shown together with the best fit model (right panel).

%--------------------------------------------------------------------
\section{Results}
\label{Results}

The analysis of the AGILE data has important implications for both the model and the investigation of diffuse \gray emission. Concerning the former, a significant part of the emission from the ROI is expected to come from molecular hydrogen. The comparison among the modelled and the measured \gray fluxes provides us with an interesting method for the estimation of the X$_{\textrm{CO}}$ factor. Also, the detection of localised flux excess can be used to reveal regions where the model fails, hinting at presence of dark gas, or a non-linear relation between H$_2$ and CO densities, or, alternatively, overabundances of CRs.

\subsection{Estimation of the X$_{\textrm{CO}}$ factor}
The calculation of the best fit proportionality factors $\alpha$ and $\beta$ allows for an estimation of the conversion factor X$_{\textrm{CO}}$ by comparing the contributions of $S_{\textrm{HI}}$ and $S_{\textrm{H}_2}$ to the total flux. Assuming that both $n_{\textrm{HI}}$ and $n_{\textrm{H}_2}$ were correctly calculated, we would expect $\alpha$ and $\beta$ to have the same value. A difference between the two values, instead, would naturally be attributed to the highest source of uncertainty in the column density estimators, namely the conversion factor X$_{\textrm{CO}}$. The effective conversion factor, (X$_{\textrm{CO}}$)$_{\textrm{eff}}$, could then be calculated as 
\begin{equation}
(\textrm{X}_{\textrm{CO}})_{\textrm{eff}}= \frac{\beta}{\alpha} \, \textrm{X}_{\textrm{CO}}
\end{equation}
This computation method leads to (X$_{\textrm{CO}}$)$_{\textrm{eff}}=(1.32\pm0.05) \times10^{20}$ cm$^{-2}$ K$^{-1}$ km$^{-1}$ s,
where the error refers to the statistical uncertainty in the computation of $\alpha$ and $\beta$. It should be noted, though, that larger uncertainties may be introduced in the (X$_{\textrm{CO}}$)$_{\textrm{eff}}$ calculation by approximations in the templates of the HI and H$_{2}$ contributions to the \gray emission, $S_{\textrm{HI}}$ and $S_{\textrm{H}_{2}}$. 
These could cause a systematic error up to $30\%$ in the estimation of (X$_{\textrm{CO}}$)$_{\textrm{eff}}$.

The $(\textrm{X}_{\textrm{CO}})_{\textrm{eff}}$ value we obtained is in excellent agreement with previous measurement provided by $Fermi$ and EGRET, respectively ($1.36\pm0.02) \times10^{20}$ (see \citealt{Ackermann12}) and ($1.35\pm0.15) \times10^{20}$ cm$^{-2}$ K$^{-1}$ km$^{-1}$ s \citep{Digel99}.

\subsection{Excess of \gray emission around ({\it l, b)}=[214.4, -18.5]}

The best fit model from Eq. \ref{eq:fit} seems to fit the AGILE data very well. After removing the isotropic extragalactic component, most of the detected \gray diffuse emission in the map can be attributed to the molecular hydrogen clouds in the region of Orion, with a fainter contribution coming from atomic hydrogen. However, a careful examination of the residuals from the fitted model reveals some excess of emission not accounted for by our template (see Fig. \ref{fig:ring}).

The most prominent feature is an overabundance of \gray photons in Region I of Orion A. Most of the emission is concentrated in a moderately extended spot whose approximate position is ({\it l, b)}=[213.9, -19.5]. This result confirms a similar analysis carried out by \cite{Ackermann12} on {\it FERMI} LAT data, showing that, also at the lower tail of energy where AGILE is most sensitive, the currently available atomic and molecular hydrogen distributions cannot fully explain the whole diffuse emission in the Orion region. 

We performed likelihood analyses of the detected excess of gamma-ray flux under the assumptions of both extended and point-like emission. In the hypothesis of extended emission, the detection is significant at a $5.2\sigma$ level, with an estimated flux of $11\pm2 \times 10^{-8}$ photons cm$^{-2}$ s$^{-1}$, while in the case of a point-like source, the analysis returns a detection significance at a $4.4\sigma$ level, and an estimated flux of $6.8\pm1.6 \times 10^{-8}$ photons cm$^{-2}$ s$^{-1}$. These results, which moderately favour the hypothesis of extended emission, do not allow a certain identification of its origin; this is not surprising, given the fact that the PSF of AGILE is almost as large as the analysed feature. In this respect, it should be noted that the detection of a point-like feature in our maps doesn't imply that the source of the emission is point-like. The assumption of an extended source of emission, nevertheless, is strongly supported by several arguments: the {\it FERMI} analysis of the same spot pointed towards an extended emission too; there are no counterparts at different wavelengths to support the idea of a point-like source. Finally, it appears that the excess-emission discussed above is not isolated; it seems to be part of an extended, continuous region approximately shaped as an arc. 

To assess the importance of the arc-shaped feature, which could hardly be compatible with a point-like source, with respect to analogous extended structures that may be present in the map of residuals, we developed a script that integrates the residual flux over circles of variable size, centred on any point of the ROI. For the whole set of integrated fluxes it calculates average and standard deviation. The only feature with integrated flux reaching approximately a 5$\sigma$ detection is a circle with $1^\circ$ radius centred in ({\it l, b)}=[214.4, -18.5] (see the white circle in Fig. \ref{fig:ring}). The nature of this emission (which, according to the arguments above, from now on will be assumed to be extended) will be thoroughly discussed in Sec. \ref{Discussion}. A further source of \gray excess, less significant, can be found at the centre of this circle.

\subsection{Unidentified source in ({\it l, b)}=[215.8, -11.2]}
\begin{figure}
   \centering
   \includegraphics[width=0.96\columnwidth]{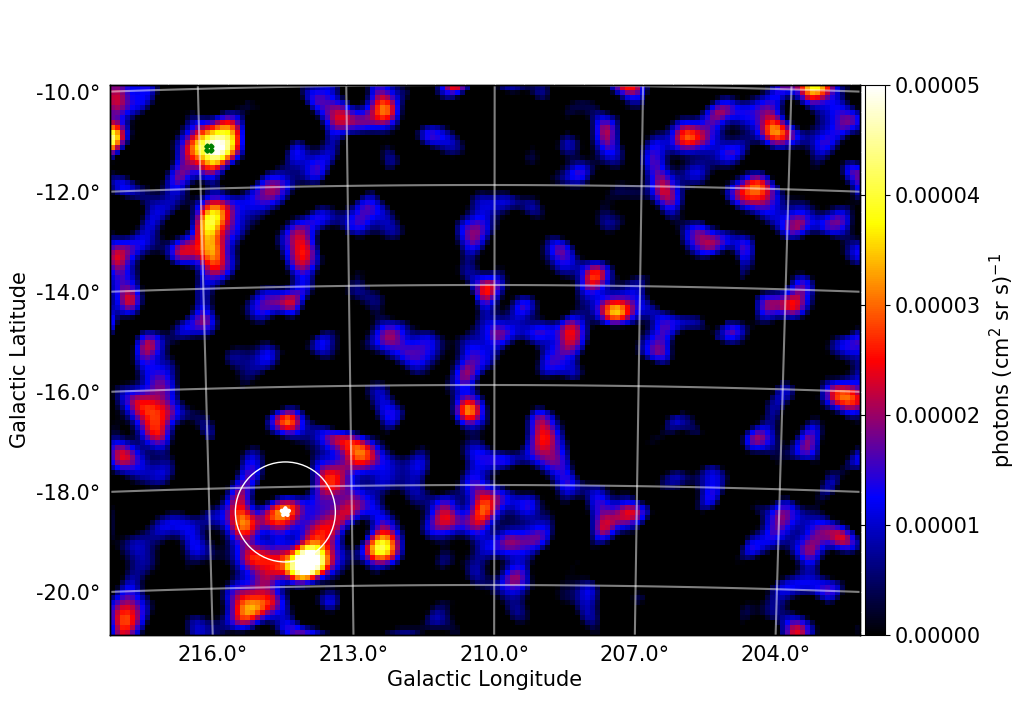}
      \caption{
      Residuals of the best fit diffuse emission model to the AGILE map. The white circle shows the ring of \gray-photon excess centred on $\kappa$ Orionis (white star). The green dot on the top-left part of the map highlights the excess of emission corresponding to the unidentified EGRET source 3EG J0616-0720.
	}   
      \label{fig:ring}
\end{figure}

Going back to the map of residuals, another interesting feature is the excess of emission around ({\it l, b)}=[215.8, -11.2] (see green dot in Fig. \ref{fig:ring}), which is significant to a $3.9\sigma$ level, with an estimated flux of $4.4\pm1.2 \times 10^{-8}$ photons cm$^{-2}$ s$^{-1}$. Most likely, it can be associated to the EGRET source 3EG J0616-0720, classified as unidentified object, whose coordinates in the Third EGRET Catalog \citep{Hartman99} are ({\it l, b)}=[215.58, -11.06], with an uncertainty in the position of $0.6-0.9$ degrees. It should be noted that 3EG J0616-0720 is among the 107 sources that have not been inserted into the EGret Revised (EGR) source list \citep{Casandjian08} because of additional structure in the interstellar background. This leaves open the question whether the emission should be associated to a point-like source or to extended structures in the ISM.

%--------------------------------------------------------------------
\section{Discussion}
\label{Discussion}

\subsection{Integrating dark gas tracers into the model}
Following \cite{Grenier05}, local excess of \gray diffuse emission can be interpreted as the result of the interaction between cosmic rays and clouds of cold dust and dark gas that are not traced by atomic hydrogen and CO surveys. We assessed the possibility of improving our fit of the \gray diffuse emission detected by AGILE by integrating dark gas tracers into a new model, which can be expressed as
\begin{equation}
S_{\textrm{tot}}(l, b) = \alpha \,S_{\textrm{HI}}(l, b)+\beta \,S_{\textrm{H}_2}(l, b)+ \gamma \,S_{\textrm{DG}}(l, b)+\epsilon
\label{eq:fit2}
\end{equation}
where $S_{\textrm{DG}}(l, b)$ is an estimate of \gray radiation from the interaction of dark gas with cosmic rays. 

\begin{figure*}
   \centering
   \includegraphics[width=0.98\textwidth]{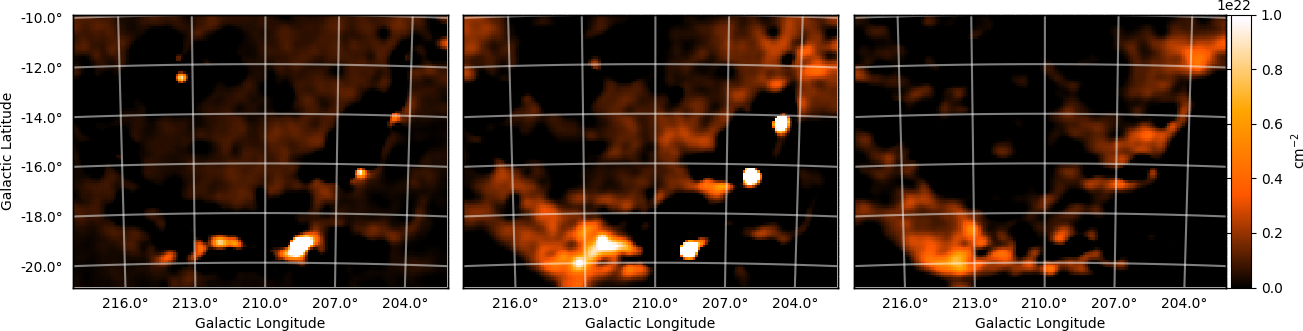}
      \caption{
      Column densities of dark gas estimated from the Planck 353 GHz map (left panel), the reddening map by \cite{Schlegel98} (middle panel), and the one by \cite{Schlafly14} (right panel).
	}   
      \label{fig:dg}
\end{figure*}

For the calculation of $S_{\textrm{DG}}(l, b)$ three different templates of dark gas have been created, which use as starting points the Planck 353 GHz map (from now on, tracer A), the reddening map derived by IRAS and COBE data by \cite{Schlegel98} (tracer B), and the reddening map derived from Pan-STARRS1 stellar photometry by \cite{Schlafly14} (tracer C). For all three maps, the procedure followed for the estimation of dark gas clouds in the region of Orion is the same: 
\begin{itemize}
\item column densities have been derived from the original maps. The procedure for the conversion of the Planck map follows the approach of \cite{Kauffmann08}. For the extinction-to-column density conversion, we applied instead the conversion factor from \cite{Guever09}.
\item Atomic and CO-traced molecular hydrogen contributions have been subtracted from the column densities. Note that, given some degree of uncertainty in the conversion of the original data into column densities, proportionality factors have been applied to the HI and H$_2$ contributions in order to minimise the residuals of the subtraction. This means that the degree of correlation between the spatial distributions of HI/H$_2$ and dark gas is minimised too.
\item The residuals of the subtraction have been retained as templates of dark gas.
\end{itemize}
The choice of minimising the correlation between the spatial distributions of HI/H$_2$ and dark gas may have some minor effect on the estimation of the best fit parameters of the model, in the sense that part of the contribution due to dark gas, e.g., could be accounted for as a H$_2$ contribution. This, however, should not affect the calculation of residuals of the new modelling.

The three tracers of dark gas are shown in Fig. \ref{fig:dg}. While tracer A shows very localised regions of dark gas (left panel), with close to no correlation with the \gray excess in Fig. \ref{fig:ring}, the tracers inferred from reddening maps suggest the existence of an extended distribution of dark gas in Region I of Orion A. In the case of tracer C (right panel), the distribution of dark gas is mainly concentrated around a ring approximately centred in ({\it l, b)}=[214.4, -18.5], which makes it the most interesting candidate for explaining the \gray excess detected by AGILE.

\begin{figure}
   \centering
   \includegraphics[width=0.96\columnwidth]{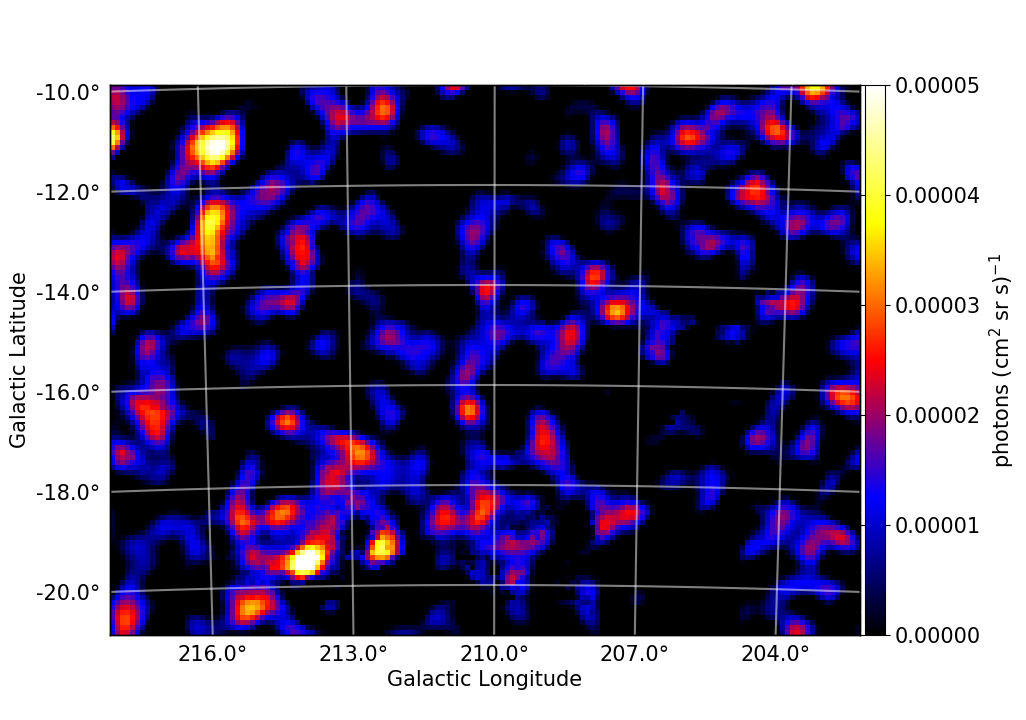}
      \caption{
      Residuals of the best fit diffuse emission model to the AGILE map, after including the dark gas tracer C into the calculation.
	}   
      \label{fig:ring2}
\end{figure}

We therefore used tracer C as a template for $S_{\textrm{DG}}(l, b)$, and recalculated the best fit model from Eq. \ref{eq:fit2}. The residuals of the improved modelling are shown in Fig. \ref{fig:ring2}. It may be surprising to see that the similarities between the dark gas tracer and the AGILE \gray excess do not translate into a substantial reduction of the excess flux. The only noticeable difference with respect to the model expressed in Eq. \ref{eq:fit} is a $\sim$25\% decrease of the residual of \gray flux in ({\it l, b)}=[213.9, -19.5]. The relatively low impact of the improved model on the fit can be explained with some significant difference in the distribution of the dark gas with respect to the diffuse \gray emission. Both dark gas and \gray estimates point towards the existence of a ring around ({\it l, b)}=[214.4, -18.5]; the diameter of the ring and the distribution of the material/emission along the ring, though, are not the same for the two cases. Apparently, either our dark gas tracers are not able to detect all the medium in Region I of Orion A, or the \gray excess is not due to dark gas, but rather to a localised overabundance of cosmic rays.

\subsection{$\kappa$ Orionis ring}

As mentioned above, the ring-shaped excess of \gray photons in Region I of Orion A has approximately the same centre ({\it l, b)}=[214.4, -18.5] as the ring in our dark gas tracer C. Considering the approximation due to the smoothing of the AGILE map and the uncertainty in the reconstruction of the origin of gamma rays, the centre of the ring is compatible with the position of $\kappa$ Orionis, ({\it l, b)}=[214.5, -18.5]. This is a blue supergiant star of spectral type B0.5 Ia, which parallax measurements place at a distance of about 200 pc (see \citealt{vanLeeuwen07}). There is evidence supporting the hypothesis that this association is not a pure coincidence. A recent work by \cite{Pillitteri16} found strong hints of the existence of a star-forming ring of 5-8 pc radius around $\kappa$ Orionis, which the authors identified by looking at the X-ray luminosity function of the young stellar objects (YSO) detected in two fields between L1641 S and $\kappa$ Orionis. They hypothesised that the ring is part of a shell in which the gas and the dust from the Orion-Eridanus superbubble have been swept up by the strong winds rising from the star, triggering star formation at the edge of the shell. The same shell has been reported in near-IR extinction maps from the Two Micron All Sky Survey (2MASS, see \citealt{Kleinmann94}; \citealt{Lombardi11}); it is also visible in the CO maps from \cite{Dame01}, and all ESA {\it Planck} maps between 217 and 857 GHz.

\begin{figure}
   \centering
   \includegraphics[width=0.96\columnwidth]{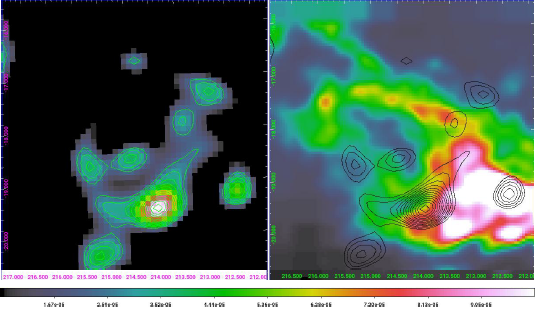}
      \caption{
      Left panel: the ring of \gray excess detected by AGILE. Right panel: the CO map from \cite{Dame01}, which reveals the star forming shell discussed by \cite{Pillitteri16}; the contour levels from the \gray data are shown in black.
	}   
      \label{fig:gammaCO}
\end{figure}

The \gray ring, with a radius of about 4 pc, could therefore be seen as the inner border of the star forming shell revealed by \cite{Pillitteri16}. This interpretation is supported by the comparison between the AGILE map and the CO emission from \cite{Dame01} shown in Fig. \ref{fig:gammaCO}. 

The existence of two concentric shells could be explained within coherent physical frameworks. Two competing scenarios seem plausible:
\begin{itemize}
\item In terms of dark gas, the winds rising from $\kappa$ Orionis could push the surrounding gas and dust towards the outside; as the temperature decreases and the density of the medium increases with distance from the star, molecular hydrogen could be formed. On the inner edge, the UV radiation that the star emits would prevent the formation of CO, which therefore would no longer trace the molecular hydrogen shell. Its interaction with cosmic rays would though reveal it as an excess of \gray photons. Farther from the star, the temperature and density of the medium, together with the lower flux of UV radiation, could provide the conditions for the formation of molecular clouds, in which star formation could be triggered. 
\item Locally accelerated CRs could provide us with a different interpretation. It involves the particle acceleration process hypothesised by \cite{Casse80} for the shocked regions where stellar wind interacts with the ISM. The acceleration of CRs in this site would naturally translate into an increase of \gray emission. Interestingly, the authors suggest Orion as an ideal candidate for testing their hypothesis, because of the high flux expected from the region. We calculated the presumable \gray flux from around $\kappa$ Orionis using the same model the authors proposed for $\rho$ Ophiuchi: F$_\gamma (\geq \textrm{100 MeV}) = 10^{-6}\, \textrm{M}\epsilon/\textrm{d}^2$ photons cm$^{-2}$ s$^{-1}$, where M is the total mass of the dense cloud in M$_\odot$, $\epsilon$ is the local \gray emissivity in 10$^{-25}$ photons ($\geq \textrm{100 MeV}$) s$^{-1}$ (H-atom)$^{-1}$, and d is the distance to the source in pc. Through the Planck function we converted the Planck 857 GHz flux from within the 4pc ring around $\kappa$ Orionis into an estimate of the mass M of the cloud, obtaining a value of $1-1.5 \times 10^3$ M$_\odot$. We estimated $\epsilon$ to be approximately 3 and, using for $\kappa$ Orionis a distance of 200 pc, we could calculate an F$_\gamma$ of the order of 10$^{-7}$ photons cm$^{-2}$ s$^{-1}$, which is consistent with the flux estimate from our AGILE data. Note that, according to \cite{Voelk82}, stellar winds would primarily be sources of nuclear cosmic rays; electron acceleration would be less favoured, which would explain why, looking at VLA maps, we didn't find any radio counterpart for the \gray emission.
\end{itemize}

The analysis of the energy spectrum from the $\kappa$ Orionis ring reveals a spectral index of $1.7\pm0.2$, indicating that the emission is hard. This suggests that, as in the case of the Cygnus Superbubble observed by $Fermi$ \citep{Ackermann11}, the origin of the radiation could more likely be a population of freshly accelerated cosmic rays. In our case, however, the acceleration would not be the result of the interaction between the stellar winds from a number of young stellar objects; it would rise as the consequence of the collision of the strong stellar wind from a single star with the surrounding ISM. The accurate modelling of this process is still on going, and will be presented in a future publication.

\subsection{$\kappa$ Orionis point-like emission}

Concerning $\kappa$ Orionis, another feature that deserves some attention is the mild excess of \gray emission within the ring surrounding $\kappa$ Orionis. It belongs to a point-like source whose position coincides with that of the B0.5 Ia star. A point-like likelihood analysis applied to the source returned a $4\sigma$ detection with a flux of $6.2\pm1.6 \times 10^{-8}$ photons cm$^{-2}$ s$^{-1}$. The origin of this \gray emission is not certain yet. The existence of a link with $\kappa$ Orionis, given the coincidence between the positions of the \gray source and the star, would be an intriguing speculation. In a theoretical work about IC scattering on ISRF around OB stars, \cite{Orlando06} indicated in $\kappa$ Orionis the most likely candidate for which \gray emission due to IC scattering could be detected by {\it Fermi}-LAT with at least 1 year of integration time. It could be hypothesised, therefore, that the origin of the \gray excess detected by AGILE is cosmic-ray IC scattering on the photons of the star. According to the model by \cite{Orlando06}, however, the emitted flux should be of the order of $10^{-9}$ photons cm$^{-2}$ s$^{-1}$ --- more than one order of magnitude lower than the one we detected. It is true that the flux values provided in the paper, according to the authors, are rather conservative; however, to justify the observed discrepancy we should suppose a substantial difference between the modelled and the actual electron spectrum in the vicinity of $\kappa$ Orionis.

%--------------------------------------------------------------------
\section{Summary}

We analysed the diffuse \gray emission collected by AGILE in the region of Orion. The data, covering the time span that goes from November 2009 to March 2017, were modelled using a template that takes into account proton-proton scattering and Bremsstrahlung emission for atomic and molecular hydrogen, while the contribution from both IC scattering on the ISRF and extragalactic background were approximated as an isotropic contribution. The model, which is characterised by three free parameters, proved to be efficient in the description of the \gray flux emitted in the Orion region. By looking at the best-fit values of the free parameters, we calculated a X$_{\textrm{CO}}$ conversion factor of ($1.32\pm0.05) \times10^{20}$ cm$^{-2}$ K$^{-1}$ km$^{-1}$ s, well consistent with previous estimations by $Fermi$ and EGRET.

The fit of the model to AGILE data allowed us to identify local spots where the \gray emission is higher than expected from the identified ISM content. A putative detection of \gray excess from the B0.5 Ia supergiant star $\kappa$ Orionis raises the intriguing hypothesis that we are observing the emission due to IC scattering on the ISRF around the star. The most interesting feature emerging from our analysis, however, is an arc-shaped excess, which appears to belong to a ring centred on $\kappa$ Orionis itself. This confirms a previous detection of \gray excess from the Orion A region reported by the $Fermi$ collaboration \citep{Ackermann12}. We hypothesise that the ring, which has a radius of $\sim4$ pc, corresponds to the inner side of a larger ring, identified by \cite{Pillitteri16} as a star forming region associated with a shell of dense gas surrounding $\kappa$ Orionis. The excess of \gray emission around the star could be due to dark gas at the edge of the shell swept up by the star, or to CR acceleration in shocked regions where the stellar wind collides with the ISM. The hardness of the energy spectrum from the \gray ring points towards the latter hypothesis. A thorougher analysis of the available data, including observations at different wavelengths, is still on going; it will hopefully help us to accurately model the acceleration process triggered by stellar wind. If our hypothesis will be confirmed, the $\kappa$ Orionis ring could represent the first direct detection of diffuse \gray emission caused by the interaction between the winds of a single OB star and the ISM.

\begin{acknowledgements}
We wish to thank the anonymous referee for the useful comments, which significantly improved the paper. We would like to thank Dr. Kazi Rygl for the stimulating discussion and the helpful suggestions. NM's research activity for this work was supported by the VIALACTEA Project, a Collaborative Project under Framework Programme 7 of the European Union funded under Contract \#607380 that is hereby acknowledged. AGILE is an ASI space mission developed with programmatic support by INAF and INFN. This study was carried out with partial support through the ASI grant no.  I/028/12/4.
\end{acknowledgements}


\bibliographystyle{aa} % style aa.bst
\bibliography{orion.bib} % your references Yourfile.bib

\end{document}